\documentclass[12pt, letterpaper, twocolumn]{article}
\usepackage{osajnl2}
\usepackage{hyperref}

\usepackage{graphicx}
\usepackage{amssymb,amsmath}

\usepackage{pstricks}
\usepackage{subfig}

\newcommand{\apjl}{{ApJL~}}        

\newcommand{\vect}[1]{\boldsymbol{\mathrm{#1}}}

\begin{document}

\twocolumn[ 

\title{Scattering from rough thin films: DDA-simulations}

\author{Hannu Parviainen$^*$ and Kari Lumme}
\address{Observatory, T\"ahtitorninm\"aki (PO Box 14), 
00014 University of Helsinki, 
Finland}

\address{$^*$Corresponding author: hannu@astro.helsinki.fi}

\begin{abstract}
We investigate the wave-optical light scattering properties of deformed thin circular films of
constant thickness using the discrete-dipole approximation. 
Effects on the intensity distribution of the scattered light due to different statistical roughness
models, model dependant roughness parameters, and uncorrelated random small-scale porosity of the
inhomogeneous medium are studied. 
The usability of discrete-dipole approximation to rough-surface scattering problems is evaluated
by considering thin films as computationally feasible rough-surface analogs. 
The effects due to small-scale inhomogeneity of the scattering medium are compared with the analytic
approximation by Maxwell Garnett and the results are found to agree with the approximation.
\end{abstract}

\maketitle

]

\section{Introduction}
\label{sec:introduction}
The discrete-dipole approximation (DDA) \cite{Purcell1973,Draine1988} is a flexible method for
calculating wave-optical interaction between electromagnetic radiation and object of arbitrary
shape. In the DDA, the scattering object is discretized into a three-dimensional lattice of dipoles, 
whose interaction with the incident radiation and with each other is then solved.

One of the main advantages of the DDA method is the flexibility to study not only objects of complex
geometries but also of complex composition. Computation of the full wave-optical solution on the
interaction between electromagnetic radiation and inhomogeneous medium is something that only a few
numerical methods are capable of, and very important when studying the limits of applicability of
different approximative scattering solutions.

Since the DDA is computationally demanding, memory requirements increasing rapidly with the increasing 
size parameter of the object, the DDA has been predominantly used to compute the scattering and
absorption properties of small particles, such as interstellar grains
\cite{Purcell1973,Draine1988,Wright1989,Lumme1994}.
Recently, the advances in computing power and numerical methods ---such as the distribution of the
geometry to different computing nodes as in the ADDA code \cite{Yurkin2007}---has allowed the DDA
simulations of objects with extended sizes. We are approaching the range where we can study
the wave-optical effects due to the rough boundaries between extended media of different physical
properties (i.e. rough-surface scattering), together with the volume effects from the internal
structure of the media. 

This paper considers computationally light rough-surface analogs, deformed thin circular films.
Rough thin films allow us to study the wave-optical effects due to surface roughness, and to
investigate the volume scattering effects to some extent.
The structure of the paper is as follows.
Section \ref{sec:introduction} introduces the reader to the DDA and current paper.
In Sec. \ref{sec:theory} we give an overview of the theoretical basis of the DDA-method, random
fields used to define the roughness deformations and the effective medium approximation. 
A quick overview to the numerical methods used in this study is given in Sec.
\ref{sec:numerical_methods}.
In Sec. \ref{sec:simulations} we present the different sets of simulations studying the varying
aspects of the scattering problem and the set specific parameters.
The results of the simulation sets are presented and discussed in Sec. \ref{sec:results}.
Finally, in Sec. \ref{sec:discussion} we discuss the limitations of the study, possible
sources of error and future prospects.

\section{Theory}
\label{sec:theory}

\subsection{Discrete-Dipole Approximation}
\label{subsec:dda}

In the DDA the solid object is approximated by an array of $N$ point dipoles at positions $\vect{r}_j$
and polarizabilities $\vect{\alpha}_j$ \cite{Purcell1973,Draine1988,Draine1993}, with the spacing
between the dipoles small compared to the wavelength. The dipole moment $\vect{P}_j =
\vect{\alpha}_j \cdot \vect{E}_{\mathrm{ext}}(\vect{r}_j)$ of each dipole responds to the total
electric field at its position. The total electric field $\vect{E}_{\mathrm{ext}}(\vect{r}_j)$ is
the sum of an incident plane wave,
\begin{equation}
\vect{E}_{\mathrm{inc},j} = \vect{E}_0 \exp(i\vect{k} \cdot \vect{r}_j - i\omega t),
\end{equation}
and a contribution from all the other dipoles
\begin{equation}
\vect{E}_{\mathrm{self},j} = - \sum_{k \neq j} \vect{A}_{jk} \cdot \vect{P}_k.
\end{equation}
Term $- \vect{A}_{jk} \cdot \vect{P}_k$ is the contribution to the electric field at position $j$
due to the dipole at position $k$
\begin{multline}
	\vect{A}_{jk} \cdot \vect{P}_k = \frac{\exp(i k r_{jk})}{r_{jk}^3} \{ k^2 \vect{r}_{jk}
\times (\vect{r}_{jk} \times \vect{P}_k) + \\ \frac{1-ikr_{jk}}{r^2_{jk}} \left[  r^2_{jk}
\vect{P}_k - 3\vect{r}_{jk}(\vect{r}_{jk} \cdot \vect{P}_k) \right ] \} \quad (j \neq k),
\end{multline}
where $\vect{r}_{jk} \equiv \vect{r}_j - \vect{r}_k$ and $r_{jk} = |\vect{r}_{jk}|$.
The dipole moments $\vect{P}_i$ are solved for all dipoles \cite{Goodman1991} and the scattering and
absorption are then computed from the dipole moments.

\subsection{Rough Thin Films}
The film geometry is represented as a thin circular slab of constant thickness $t$ along the
$z$-axis. The deformation of the film along the $z$-axis is modeled by a two-dimensional homogeneous
isotropic random field $h(x,y)$ \cite{Adler1981,Preston1976,Ibragimov1970}. The distribution of
heights follows zero-mean Gaussian statistics and is defined by the standard deviation $\sigma$.
The field realizations are periodic in $x$ and $y$, with length of the period $L$. The generation of
the geometry realizations is based on the spectral synthesis method
\cite{Fenton90a,Parviainen2006,Dieker2002}.

\begin{figure*}
	\centerline{\includegraphics[width=\textwidth]{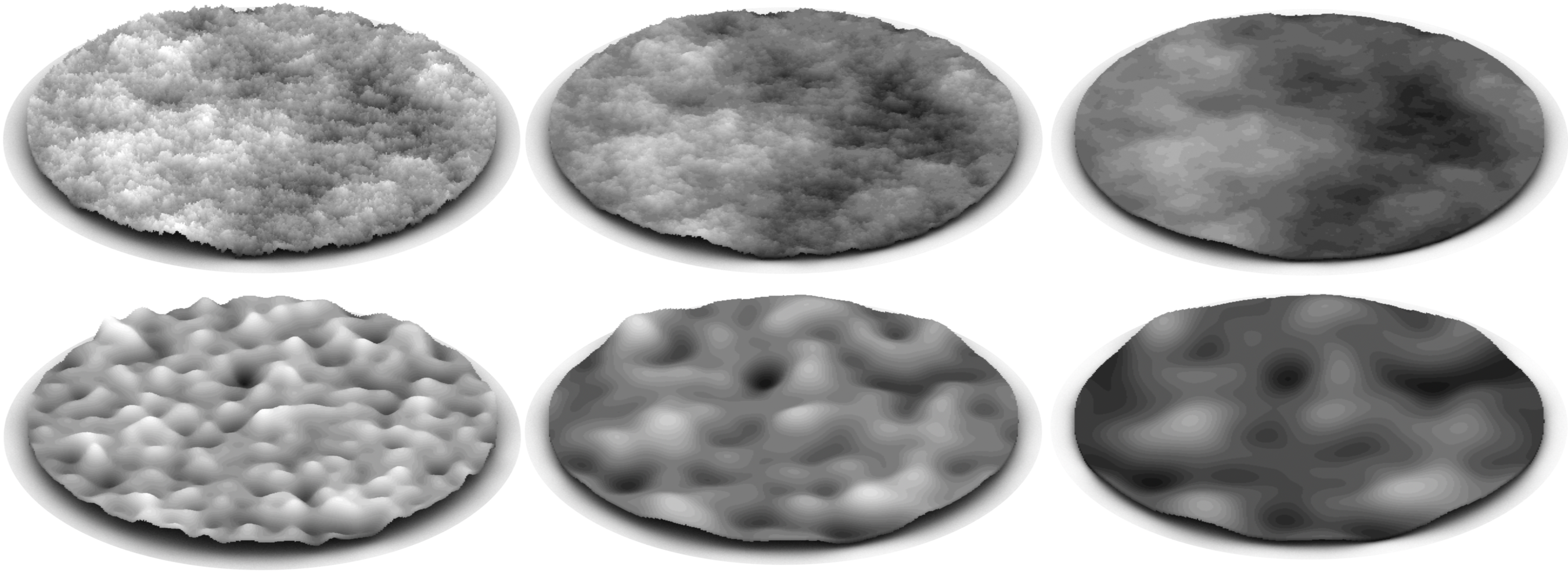}}
	\caption{ Realizations of the film geometry with fBm and Gc roughness models and horizontal
	roughness parameter $\tau$ varying from small-scale roughness to large-scale roughness. In the
	upper row we show the fBm model, in the lower the Gc model. Horizontal roughness parameter changes
	from left to right as $H = 0.25,\; 0.5,\; 0.9$ and $\frac{l}{L} = 0.215,\; 0.5,\; 0.75$.}
	\label{fig:roughness_types}
\end{figure*}

As in \cite{Parviainen2007}, two different types of roughness models were chosen for the study:
fractional Brownian motion (fBm) and Gaussian correlation (Gc). The two models are shown in Fig.
\ref{fig:roughness_types}. Gc-surfaces show roughness features of a certain scale, determined by the
correlation length $l$, while the fBm-surfaces are of self-affine nature showing roughness features
in all scales, distribution determined by the Hurst exponent $H$. For brevity, both $l$ and $H$ are
denoted in the following text by a horizontal roughness parameter $\tau$.

\subsection{Effective Medium Approximation} 
For a rough film (or surface), the scattering properties depend both on the roughness statistics and
the internal structure of the medium \cite{Guerin2007}.
When the largest scale of the fluctuations of the scattering medium inhomogeneities is smaller than
the wavelength of the incident radiation, the inhomogeneous medium can be approximated by a
homogeneous medium with effective volume-averaged properties \cite{BohrenHuffman}. The DDA-based methods
offer an intriguing opportunity to test the regions of validity of these effective medium
approximations, such as the Maxwell Garnett approximation \cite{MaxwellGarnett1904}.

The effective index of refraction $m_{\mathrm{eff}} = n_{\mathrm{eff}} + ik_{\mathrm{eff}}$ is obtained 
from the effective dielectric function $\epsilon_{\mathrm{eff}}$, which in the Maxwell Garnett 
approximation for spherical inclusions \cite{BohrenHuffman} is
\begin{equation}
	\epsilon_{\mathrm{eff}} = \epsilon_{\mathrm{m}} \left [ 1 + \frac{3f \left( \frac{\epsilon_{\mathrm{i}} -
\epsilon_{\mathrm{m}}}{\epsilon_{\mathrm{i}} + 2\epsilon_{\mathrm{m}}}\right )}{1 - f \left (  \frac{\epsilon_{\mathrm{i}} - \epsilon_{\mathrm{m}}}{\epsilon_{\mathrm{i}} +
2\epsilon_{\mathrm{m}}}\right )} \right ],
\end{equation}
where $f$ is the mixing factor of the two media, $\epsilon_{\mathrm{i}}$ the dielectric function of the inclusions and
$\epsilon_{\mathrm{m}}$ the dielectric function of the medium.

\section{Numerical Methods}
\label{sec:numerical_methods}

\subsection{DDSCAT}
\label{subsec:ddscat}
Simulations were carried out using a modified version of the DDSCAT-software \cite{Draine1988,Draine2003} by B. Draine and P. Flatau, and the results were verified against the ADDA code \cite{Yurkin2007}.
The modifications consisted of inclusion of several Fortran 90 features into the Fortran 77 code,
such as dynamical allocation of memory to allow for more convenient working with geometries
of varying sizes. The modifications were tested not to affect the simulation results.

\subsection{Computations}
The Sepeli computing cluster of the Finnish IT center for science (CSC) was used to carry out the
simulations. The memory usage for a single geometry was around 4-6 GB, and computation time varied
from several tens of minutes to hours, depending on the packing density $\rho$ and the complex
refractive index $m$ of the scattering medium.

\subsection{Rough-Film Geometry}
\label{subsec:rough_film_geometry}
The geometry was discretized into cells of 0.025 $\mu\mathrm{m}$ size. The radius of the
circular film was 5.0 $\mu\mathrm{m}$, and the thickness of the film was set to 4 cells,
corresponding to 0.1 $\mu\mathrm{m}$. This discretization scheme leads to effective radius
$A_{\mathrm{eff}}$ of 1.12, and size parameter $X = 2 \pi A_{\mathrm{eff}} / \lambda$ of 12.64 with a  value of $\lambda = 0.557 \mu m$.
Only the effects due to the horizontal roughness parameter were studied, and the amplitude of the
roughness deformations in the $z$-axis was considered as constant, $\frac{\sigma}{L} = 0.01$. 

The vertical profile of the film is determined by a two-dimensional random field, generated using
the spectral synthesis method\cite{Fenton90a,Parviainen2006,Dieker2002}. The field statistics
determine the power spectrum of the surface features in the frequency domain, and a field
realization is generated by transforming a randomized realization of the power spectrum to the
spatial domain using the fast Fourier transform (FFT).

\begin{figure*}[t]
	\centerline{\includegraphics[width=\textwidth]{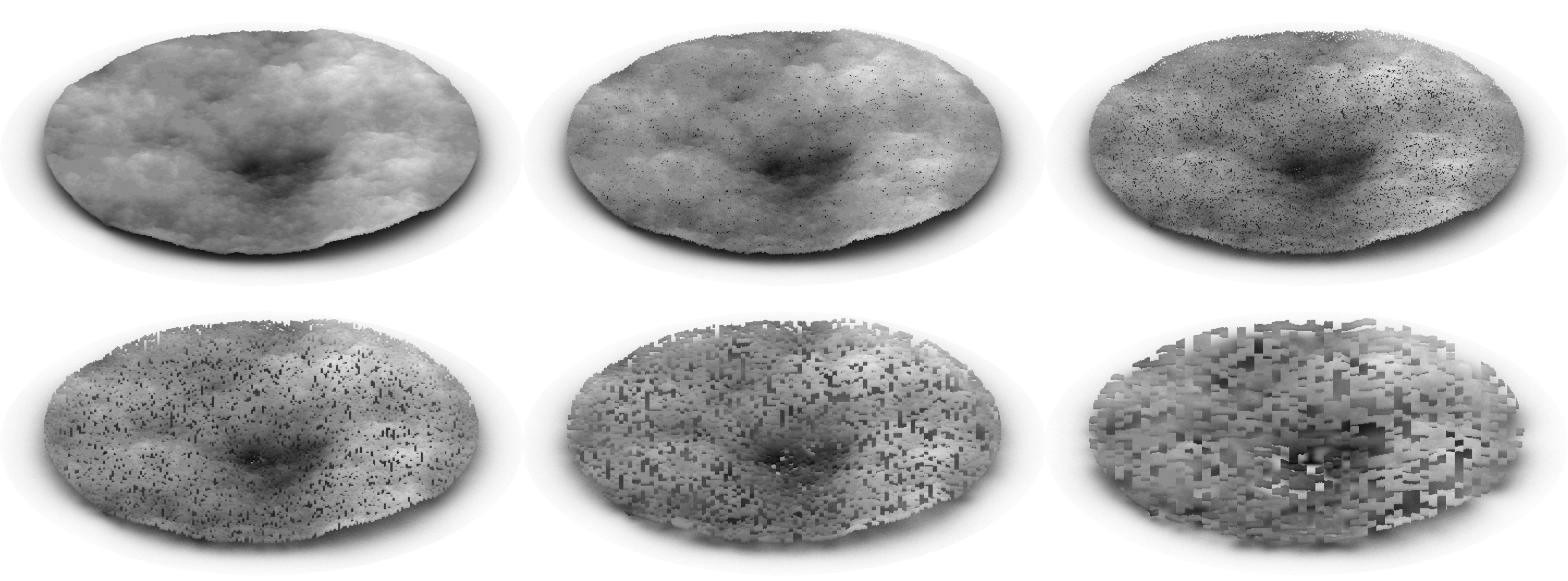}}
	\caption{ Realizations of the film geometry with different porosities and void sizes for fBm
	roughness model with $H = 0.5$. The upper row has void size of one dipole and $\rho = [1.0,\;
	0.5,\; 0.3\;]$, starting from left. The lower row has $\rho = 0.5$ and void sizes of 2, 4 and 8
	dipoles. It can be seen that even the most porous geometry with $\rho = 0.3$ still preserves most
	of its wavelength-scale ($\approx 22$ dipoles) structure.}
 \label{fig:density_types}
\end{figure*}

Geometries with packing density $\rho < 1$ were generated by randomly removing $n \times n \times n$
dipole chunks from the solid geometries, as illustrated in Fig. \ref{fig:density_types}. While this
method lacks the elegance of using, e.g., a
random field to define $\rho$, it allows us to study the scattering dependency of the films due to
the varying size of the voids in a simple and well-determined manner.

\subsection{Estimation of Accuracy}
Two general classes of error sources can be identified: the method dependent errors and
implementation dependent errors. Method dependent errors are due to the approximations taken by the
simulation method, they can often be predicted from the simulation parameters and object geometry, and
are often similar between different implementations of a single simulation method. Implementation
dependent errors are due to different numerical methods used to implement the method and possible
implementation specific bugs. Implementation dependent error sources are easy to investigate by
comparing the results from different implementations for identical geometries. The approximation of
method dependent error sources is more problematic---especially with the DDA---since there might be no
other suitable methods available to compare with.

The lack of comparable methods to compute the wave-optical solution to the inhomogeneous objects of 
arbitrary shape constrains us to use simplified geometries analogous to the simulation geometries for 
the determination of method dependent errors.

Method dependent error sources were investigated by comparing the forward scattering results from
simulation of normally irradiated undeformed homogeneous thin cylinder to the Fraunhofer diffraction
pattern by a circular aperture. The simulation results agreed well with the exact solution. This is
no real surprise, since the error sources for the DDA have been characterized by previous studies
\cite{Draine1988,Draine1993,Draine1994,Collinge2004}, and the scattering geometries were generated
to minimize the known parameters affecting the accuracy. Dipole size was small enough compared to the
wavelength of the incident radiation, and the norm of the refractive index was small. 

The implementation dependent error sources were studied by comparing the results of DDSCAT
simulations against ADDA simulations with the same geometries. The results obtained using the two
codes matched to high accuracy. 

\section{Simulations}
\label{sec:simulations}

\subsection{Illumination Geometry}
\label{subsec:illumination_geometry}
\begin{figure}
	\centering
	\includegraphics[width=\columnwidth]{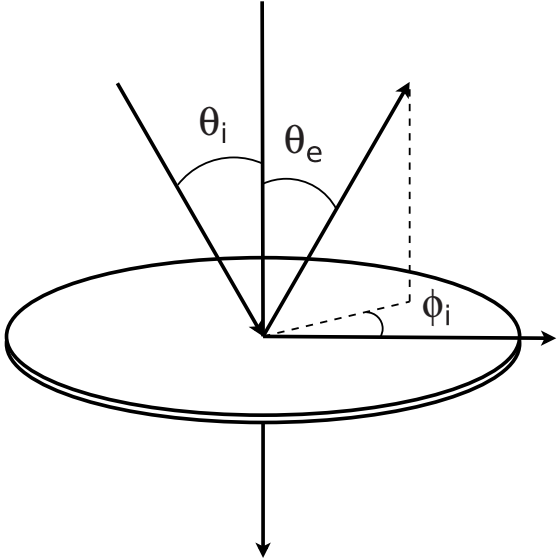}
	\caption{Simulation geometry. $\theta_i$ is the angle between the incident radiation and the
	normal of the film, $\theta_e$ between the emergent radiation and the normal of the
	film and $\phi_i$ and $\phi_e$ are the azimuth angles computed from the $x$-axis.}
	\label{fig:illumination_geometry}
\end{figure}

The main focus of the study was on the the behavior of the scattered intensity $M_{11}(\theta_i,
\phi_i, \theta_e, \phi_e)$---where $(\theta_i, \phi_i)$ are the angles of incidence, $(\theta_e,
\phi_e)$ the angles of emergence, illustrated in Fig. \ref{fig:illumination_geometry}--- 
as a function of the horizontal roughness parameter and composition of the scattering media. Two
different illumination geometries were used in the study. Most of the simulations were carried out
with the direction of the incident radiation normal to the plane of the film. To separate possible
backscattering effects from the specular scattering, simulations were also carried out with
$\theta_i = 15^{\circ}$, leading to a $30^{\circ}$ separation between the backward and specular
directions.

\subsection{Simulation Sets}
\label{subsec:simulation_sets}

The 
simulations were divided into five sets, each studying different aspects of the scattering problem.
 For each value of the studied variable, the results are
averaged over 30 geometry realizations to obtain statistically meaningful results. When possible 
(i.e., for normal incident radiation, sets 1-4), 
the averaging is also carried out over 20 values of $\phi_e$, thus giving $M_{11}(\theta_e)$ as an
average of 600 samples.

\begin{description}
\item[Set 1]
studied the behavior of $M_{11}(\theta_e)$ as a function of the horizontal roughness parameter and
density of the medium for normal incident radiation ($\theta_i = 0$). The simulations were carried
out for five values of $\tau$ ($\frac{l}{L}\in\{0.25,\; 0.3,\; 0.4,\; 0.5,\; 0.75\}$ for the Gc model,
$H\in\{0.25,\; 0.375,\; 0.5,\; 0.625,\; 0.9\}$ for the fBm model), and three values of packing density
($\rho \in \{1.0,\; 0.5,\;  0.3\}$) for a void size of a single dipole. Refractive index of the medium
$m$ was set to $1.5 + 0.001i$, yielding $|m|kd = 0.423$, which should assure relatively accurate
results for the differential quantities \cite{Draine1993}.

\item[Set 2]
considered the effects due to the imaginary part of the refraction index. Simulations were carried
out for fBm films with $\rho = 1.0$, $n = 1.5$, $k \in \{0.01i,\; 0.1i,\; 1i\}$ and $H \in \{0.5,\;
0.625\}$.

\item[Set 3]
treated the approximation of the inhomogeneous media by a homogeneous one with an average index of
refraction using the Maxwell Garnett relation \cite{MaxwellGarnett1904, BohrenHuffman}. Simulations
were carried out for fBm films with $H \in \{0.5,\; 0.625\}$, and $m$ computed using
the Maxwell Garnett relation for densities of $\rho \in \{0.5,\; 0.3\}$.

\item[Set 4]
was a follow-up study for the sets 1 and 3. The behavior of the scattered intensity was studied as
a function of the size of square-shaped voids, as illustrated in Fig. \ref{fig:density_types}.
Simulations were carried out for fBm films of $\rho = 0.5$, $H = 0.5$, with size of the voids $n \in \{2,\; 4,\; 8\}$ dipoles and two incident angles $\theta_i \in \{0^{\circ},\; 20^{\circ}\}$.

\item[Set 5]
was a follow-up study for the set 1. The simulations of the set 1 were carried out for $\theta_i =
15^{\circ}$, and no averaging over $\phi_e$ was done. This was to test the consistency of the
results obtained from the set 1, especially the behavior of the specular reflectance as a function
of varying $\rho$.
\end{description}

\section{Results}
\label{sec:results}

\begin{figure*}[t]
	\centering
	\centerline{\includegraphics[width=\textwidth]{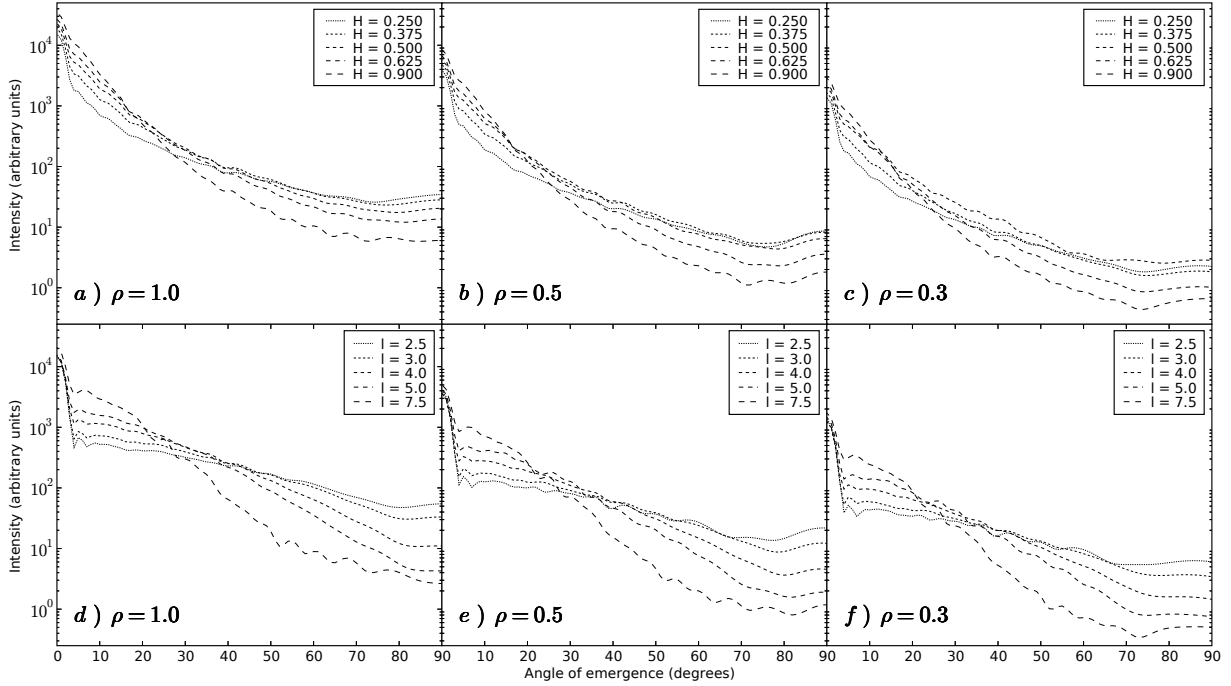}}
	\caption{
		Distribution of the scattered intensity ($M_{11}$) as a function of angle of emergence
		$\theta_e$ computed for
		the films with fBm roughness (upper row) and Gc roughness (lower row) and normal incident
radiation. The angle of emergence shown ranges from the backscattering and specular
direction $\theta_e = 0^{\circ}$ to $\theta_e =		90^{\circ}$, and the diffraction-dominated
forward-scattering direction is omitted. 
	}
	\label{fig:intensity_distribution}
\end{figure*}

\begin{figure*}[t]
	\centering
	\centerline{\includegraphics[width=\textwidth]{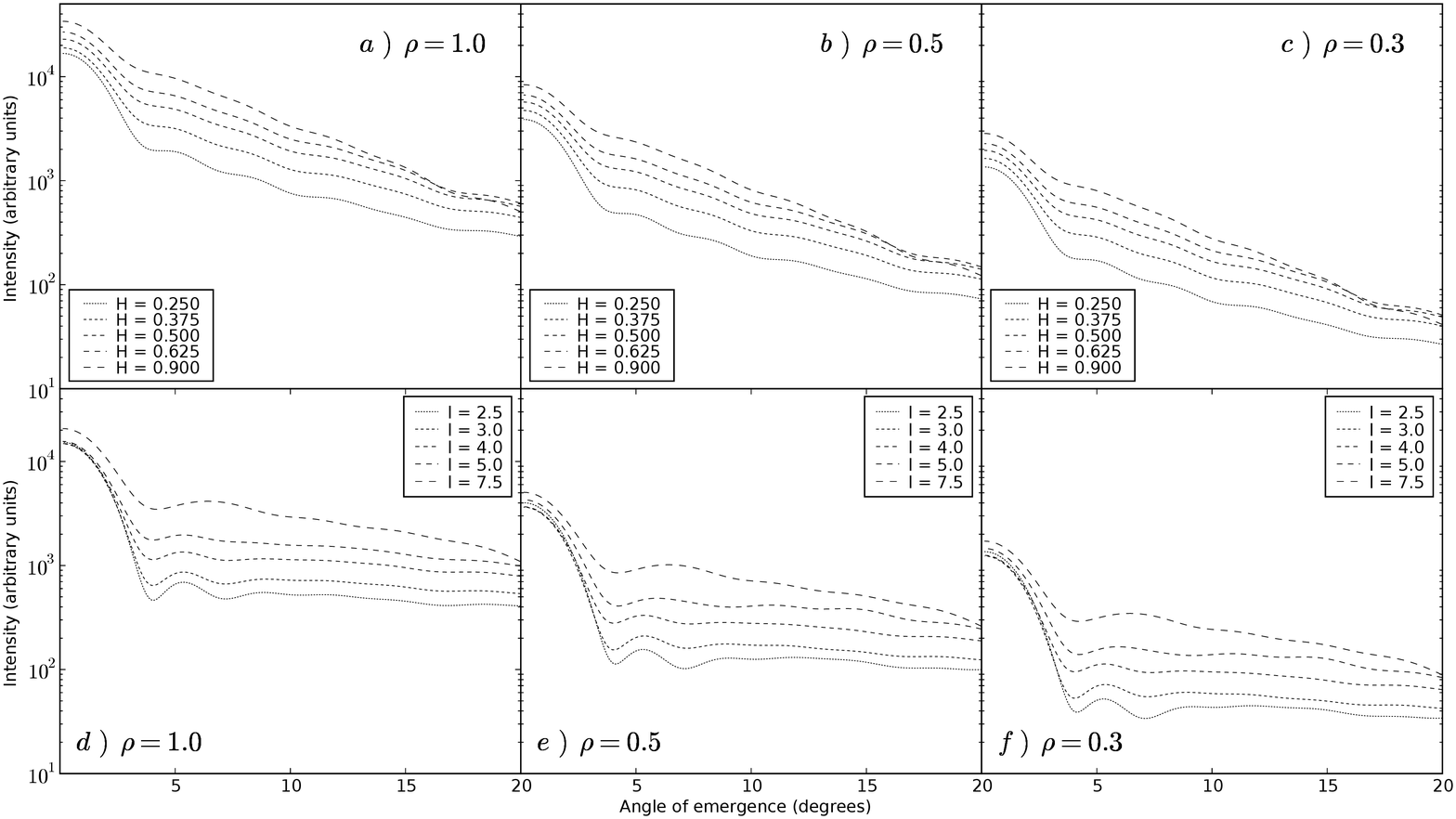}}
	\caption{As in Fig. \ref{fig:intensity_distribution} but for $\theta_e \in \{0^{\circ},20^{\circ}\}$. 
	}
	\label{fig:intensity_distribution_gloss}
\end{figure*}

\begin{figure*}[t]
	\centering
	\includegraphics[width=\textwidth]{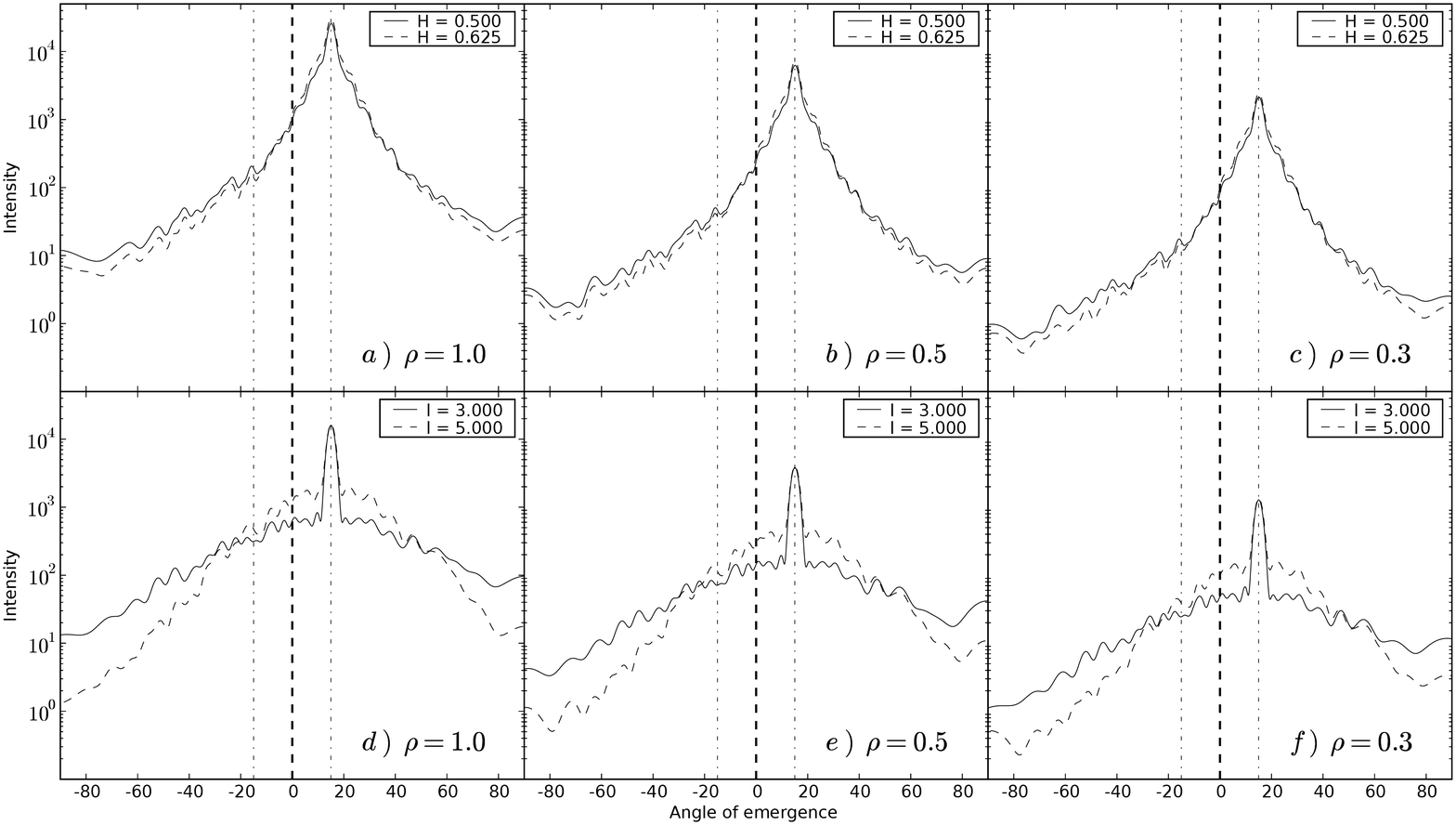}
	\caption{
		As in Fig. \ref{fig:intensity_distribution} but for $\theta_i = 15^{\circ}$,  $\theta_e =
		[-90^{\circ}..90^{\circ}]$, $H\in\{0.5,\; 0.625\}$ and $l \in \{3.0,\; 5.0\}$	The limits for the angle
		of emergence are different since the distribution is no longer symmetric around $\theta_e=
		0^{\circ}$. The shapes of the scattering distributions follow the shapes for normal incident
		radiation, with clear distinction between the fBm and Gc roughness models.
	}
	\label{fig:T15}
\end{figure*}

\begin{figure*}[t]
	\centering
	\centerline{\includegraphics[width=\textwidth]{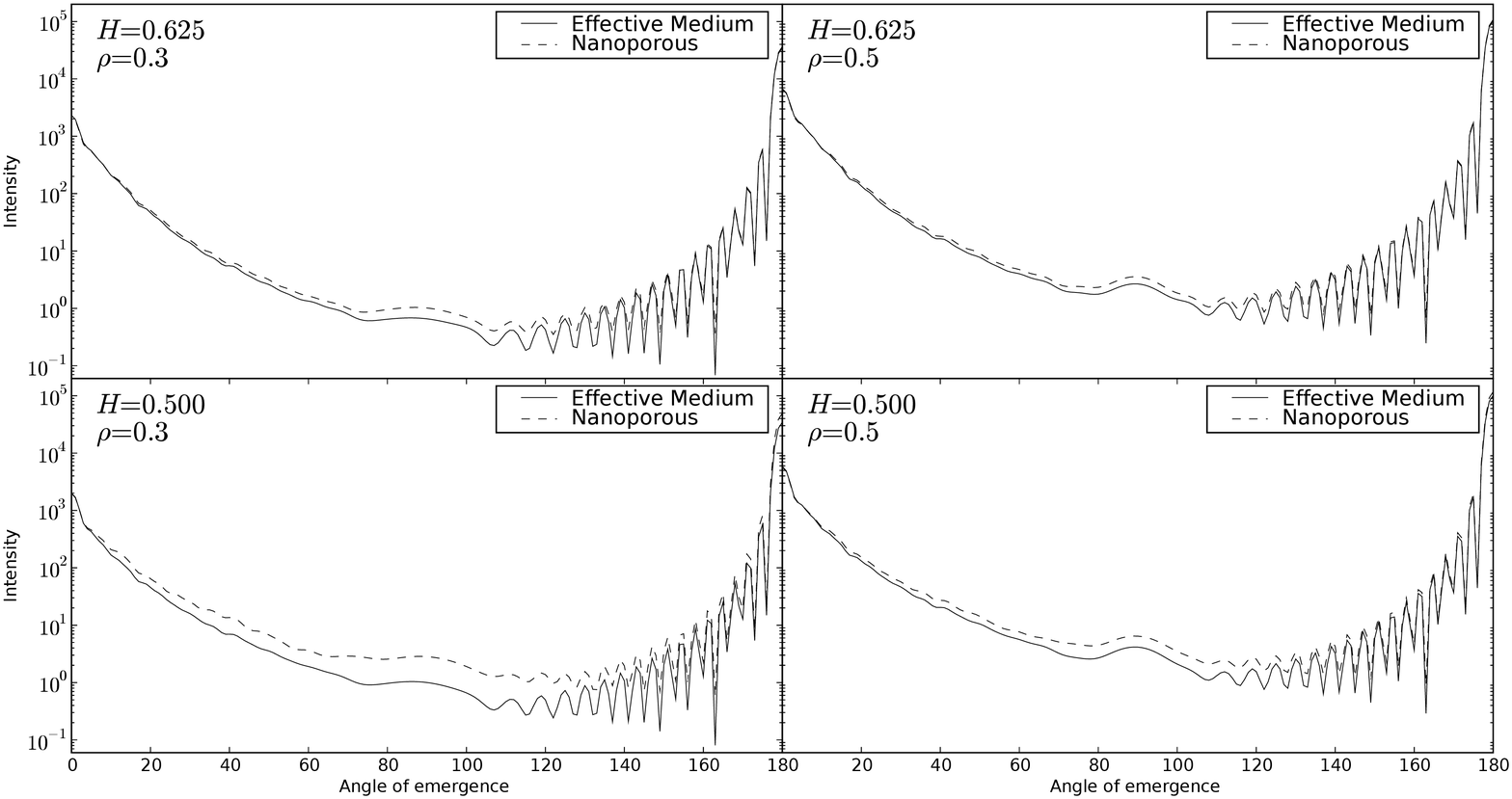}}
	\caption{
		Distribution of scattered intensity for inhomogeneous films with $\rho = [0.5,\; 0.3]$, and 
		solid films with effective index of refraction. Here we show the results for normal incident
	radiation and $\theta_e = [0^{\circ}..360^{\circ}]$ (we include the diffraction-dominated
	forward-scattering direction for clarity).}
	\label{fig:effective_medium_back}
\end{figure*}

Figures \ref{fig:intensity_distribution}, \ref{fig:intensity_distribution_gloss}, \ref{fig:T15} and
\ref{fig:effective_medium_back} sum up the primary results of the simulations. In Fig.
\ref{fig:intensity_distribution} we show the results from simulation set 1, i.e., effects due to
varying density, horizontal roughness parameter and different roughness model. Figure
\ref{fig:intensity_distribution_gloss} is a close-up of Fig. \ref{fig:intensity_distribution},
showing the specular and backscattering region $\theta_e = [0^{\circ},20^{\circ}]$ in greater
detail.
Figure \ref{fig:T15} displays the results for the simulation set 5, off-normal incident radiation.
The results are in good agreement with the results for normal incidence.

\begin{figure}[t]
	\centering
	\centerline{\includegraphics[width=\columnwidth]{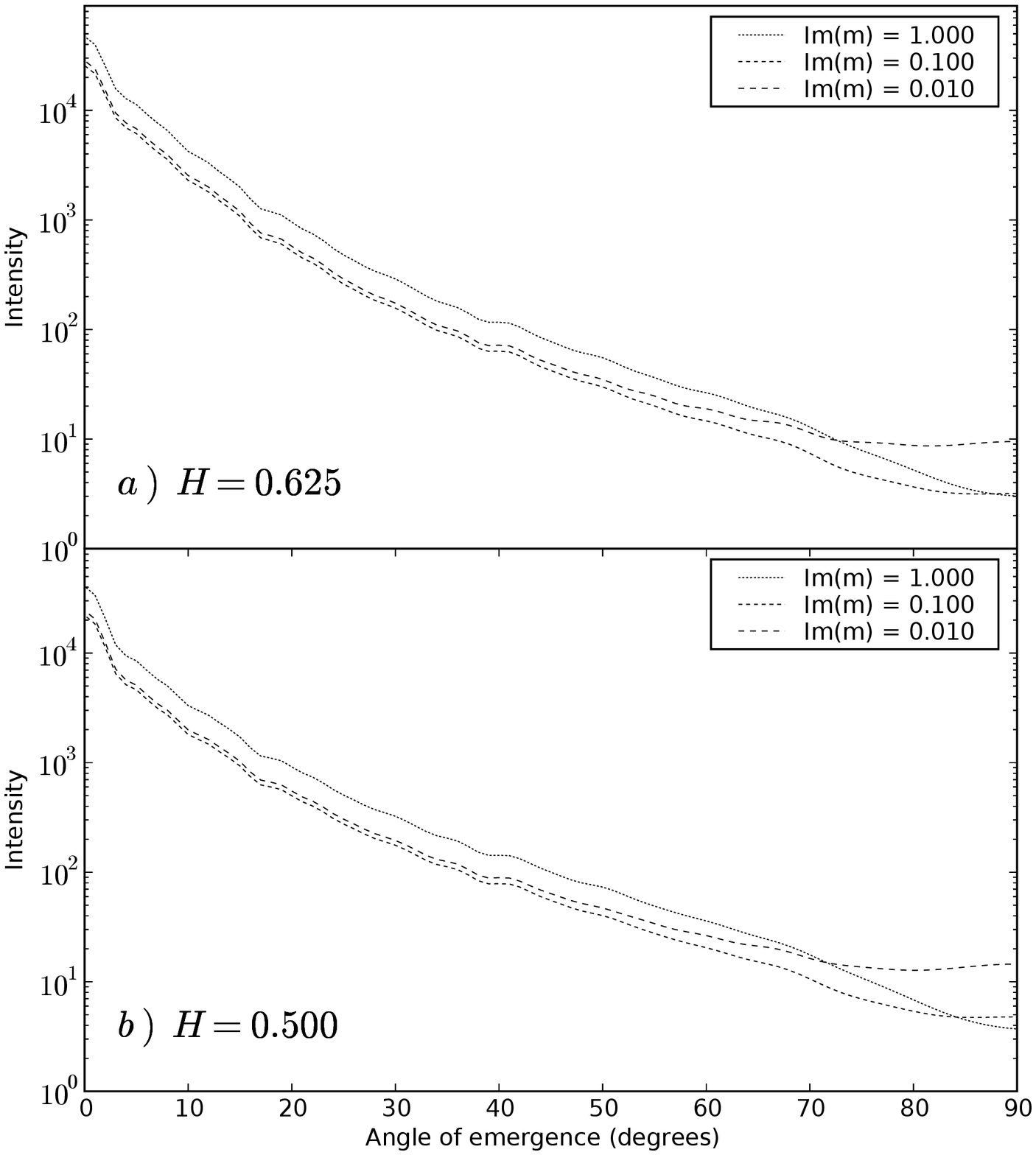}}
	\caption{Distribution of scattered intensity shown for fBm films with $H \in \{0.5,\; 0.625\}$, normal
		incident radiation	and three values for the imaginary part of the refractive index ($k \in
		\{0.01,\; 0.1,\; 1.0\}$).}
	\label{fig:index_of_refraction}
\end{figure}

In Fig. \ref{fig:index_of_refraction} we show the the effects due to varying imaginary part of the
refractive index for fBm roughness with $H \in \{0.5,\; 0.625\}$. The functional shape is nearly
constant for $\theta_e = [0^{\circ},60^{\circ}]$
and each value of $m$. The conclusion can be made that the imaginary part of the refractive index is
of little importance for the backward intensity distribution from homogeneous rough films.

The results of the simulation set 3, comparison of inhomogeneous porous films and homogeneous solid
films with effective index of refraction, is shown in Fig. \ref{fig:effective_medium_back}. Here we
show the scattering for the full range of $\theta_e$, including the forward scattering region. The
agreement between porous media and solid media with index of refraction computed using the relation
by Maxwell Garnett is very good for films with $\rho = 0.5$. Largest deviation between the two
cases is found for films with $H=0.5$ and $\rho = 0.3$, i.e., very porous films with
small-scale surface roughness. 

\begin{figure}[t]
	\centering
	\centerline{\includegraphics[width=\columnwidth]{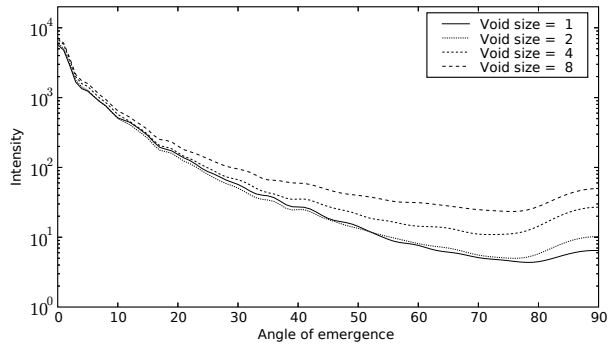}}
	\caption{Distribution of the scattered intensity from inhomogeneous films as a function of
	emergent angle $\theta_e$ for		normal incident radiation and four values for the void size. Shown
	are the results for	films with $\rho	= 0.5$ and fBm roughness of $H = 0.5$. The results for void
	size of one dipole correspond the results shown in Fig. \ref{fig:intensity_distribution}b.}
	\label{fig:blocky_voids}
\end{figure}

\begin{figure}[t]
	\centering
	\centerline{\includegraphics[width=\columnwidth]{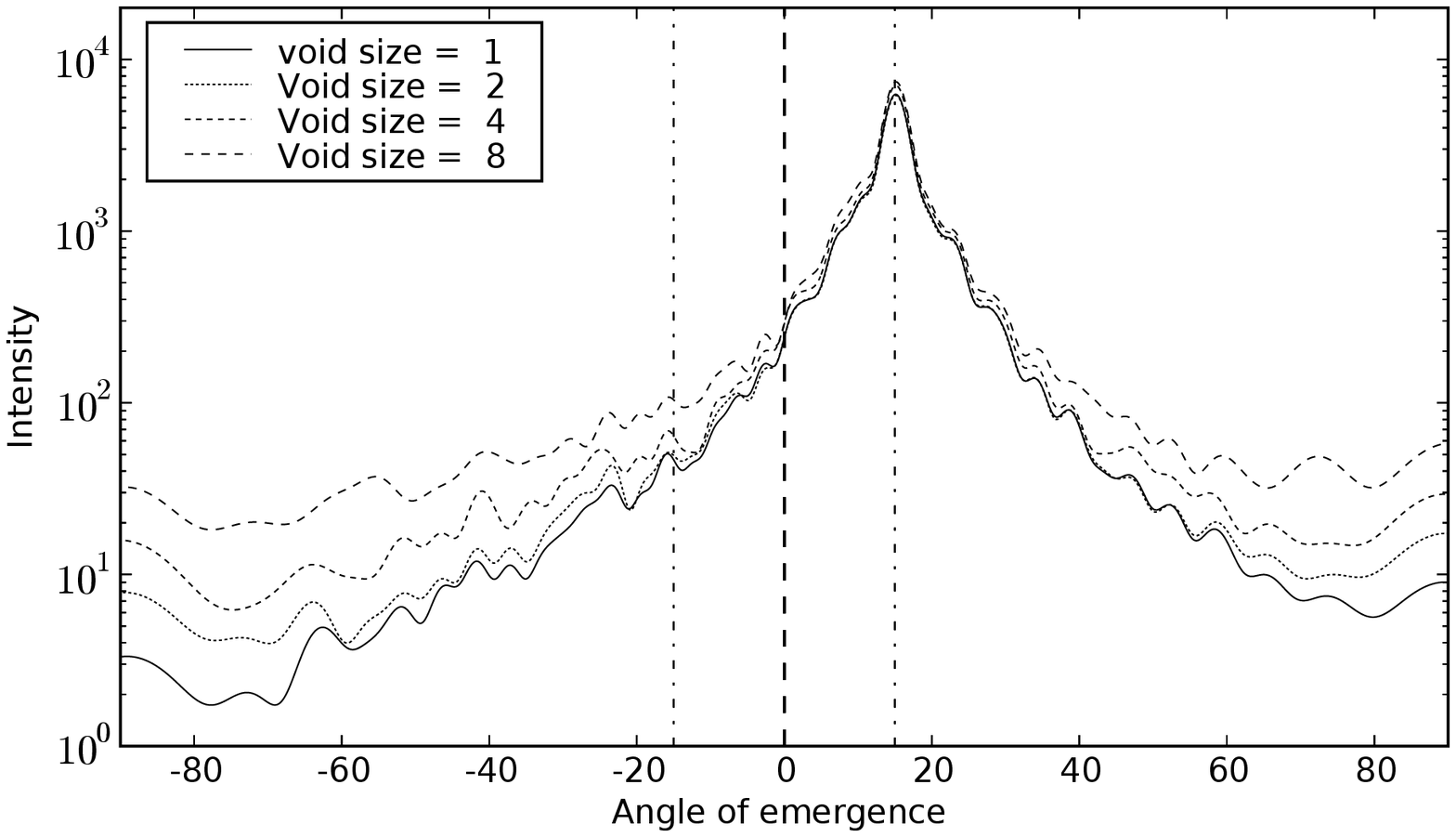}}
	\caption{As in Fig. \ref{fig:blocky_voids} but for $\theta_i = 15^{\circ}$.}
	\label{fig:blocky_voids_t15}
\end{figure}

Figures \ref{fig:blocky_voids} and \ref{fig:blocky_voids_t15} display the effects due to the
growing size of the voids for inhomogeneous films. The largest void size, sidelength of 8 dipoles,
corresponds to $\approx\frac{1}{3} \lambda$. The most notable result is the relatively small
effect to the intensity distribution near the specular direction. This can be due to the extremely
thin nature of the films.

The most prominent results are:

\begin{enumerate}
\item The Gc and the fBm models lead to rather a different distribution of the backward scattered
	intensity. For the fBm model, the transition from the specular reflection to diffuse is smoother
	than for the Gc model. For both models, the specular peak smoothens when the mean scale of the 
	roughness increases. This agrees with the theories based
	on the wave-optics: the directional diffuse component of the scattered radiation starts to dominate
	the specular reflectance when the scale of the roughness approaches the scale of the wavelength.

\item Approximation of the inhomogeneous medium using solid geometry with the relation by Maxwell
	Garnett agrees with	the simulations.
	The shape of the 	reflectance is not sensitive to the inhomogeneities 
	of scale $\frac{l}{\lambda} \approx 0.05$, even for a loose geometry with $\rho = 0.3$.
	Decreasing density is manifested as a multiplicative factor constant over $\theta_e$, with
	only minor differences in the shape of the reflectance distribution.

\item While the thinness of the geometry prevents us from studying the volume scattering effects in
	depth, basic conclusions can be made from the behavior of the reflectance as a function of void
	size. From Fig. \ref{fig:effective_medium_back} we see that the results deviate from the effective
medium 
	approximation along with the increasing void size, but the relative deviation reduces near
	the specular direction. Nevertheless, more simulations for off-normal incidence and significantly 
	increased thickness are required for a serious study of the effects due to size distribution of the 
	inhomogeneities.

\item The imaginary part of the refractive index plays only a minor role in the scattering
	distribution when constant over the whole medium. Increasing value of the imaginary part leads to
	stronger interaction between the	scattering medium and incident radiation, but does not alter the
	scattering distribution. 

\end{enumerate}

In summary, the backward scattered intensity distribution depends strongly on the roughness model
statistics, showing notably different specular behavior for the Cg and fBm models. Random
inhomogeneities of mean scale smaller than the wavelength are well approximated by effective
index of refraction, and have only a minor effect to the specular scattering.

\section{Discussion}
\label{sec:discussion}
The study considered light scattering from simple nanoporous media, that is, inhomogeneous media
with pore size in the nanometer range and simplified pore structure. The results can be generalized to
situations where the pore size is very small compared to other structures of the media---such as
the surface roughness---and to the wavelength of the radiation. The results can not be generalized
to particulate media with coherent particle or pore structure, or to porous media with pore
structure exceeding the scale of the wavelength.

The comparison of the results against implementation of the finite-difference time-domain (FDTD)
method might offer valuable information, and could be done in the future. 

For a scattering object to be considered analogous to a surface, we must have $\frac{r}{t} >> 1$.
Here $r$ is the radius of the cylinder, and $t$ the thickness. This is especially important for
off-normal incident radiation, since the cylinder walls contribute to the scattering. Nevertheless,
to include realistic volume scattering effects, we would like to have $t>>\lambda$, where $\lambda$
is the wavelength of the radiation.
With the DDSCAT, the size of the geometry is restricted by the available memory of a single computing node. This imposes a strict limit to the geometry thickness. This limit can be raised
by applying codes with capability to slice the geometry between different nodes, such as the ADDA.
Geometry slicing will allow us to maximize $\frac{r}{t_s}$ for a single computing node,
where $t_s$ is the thickness of a slice, and use appropriate number of nodes to achieve the
total thickness. 

The use of more extended rough-surface analogs will allow
for the comparison between the different analytic wave-optical approximations for random rough surfaces \cite{Elfouhaily2004} and independent numerical simulations.

\section{Conclusions}
The results from the DDA-based simulations show that the functional shape of the distribution of 
the scattering intensity $I(\theta_e)$ near the specular scattering direction is significantly different
between the two different roughness models. The refractive index and small-scale porosity of the scattering
medium have very little effect on the functional shape of $I(\theta_e)$ near the specular scattering direction.
Also, the effects due to small-scale inhomogeneity of the scattering medium agree with the analytic approximation 
by Maxwell Garnett.

\section*{Acknowledgments}
H. Parviainen thanks Antti Penttil\"a, Karri Muinonen and the anonymous referee for their constructive and helpful comments. 

This project was funded by Tekes, the Finnish Funding Agency for Technology and Innovation.
The computations were carried out using the computation facilities of the Finnish IT center for science, CSC.

\end{document}